\documentclass[useAMS,usenatbib,letterpaper]{mn2e}
\usepackage[pass]{geometry}
\usepackage{natbib}
\usepackage{amssymb,amsmath}
\usepackage{graphicx}
\usepackage{verbatim}
\usepackage{color}
\usepackage{hyperref}
\usepackage{lastpage}
\definecolor{linkcolor}{rgb}{0,0,0.25}
\hypersetup{
  colorlinks=true,        
  linkcolor=linkcolor,    
  citecolor=linkcolor,    
  filecolor=linkcolor,    
  urlcolor=linkcolor      
}
\newcommand{\etal}{et al.}
\newcommand{\dd}{\mathrm{d}}
\newcommand{\eg}{e.g.}

\newcommand{\Eqnname}{Equation}
\newcommand{\equationname}{\Eqnname}

\newcommand{\figurename}{Figure}

\newcommand{\sectionname}{$\mathsection$}

\newcommand{\kpc}{\ensuremath{\,\mathrm{kpc}}}
\newcommand{\pc}{\ensuremath{\,\mathrm{pc}}}
\newcommand{\kms}{\ensuremath{\,\mathrm{km\ s}^{-1}}}

\newcommand{\inv}{\ensuremath{^{-1}}}

\def\aj{AJ}
\def\apj{ApJ}
\def\apjl{ApJL}
\def\apjs{ApJS}
\def\mnras{MNRAS}

\def\aap{A \& A}

\def\pasp{PASP}

\title[Galactic rotation in \protect\emph{Gaia} DR1]{Galactic rotation in \protect\emph{Gaia} DR1}
\author[Jo Bovy]{Jo Bovy\thanks{E-mail: bovy@astro.utoronto.ca}\thanks{Alfred~P.~Sloan~Fellow}\\
Department of Astronomy and Astrophysics, University of Toronto, 50 St. George Street, Toronto, ON M5S 3H4, Canada\\
and\\
Center for Computational Astrophysics, Flatiron Institute, 162 5th Ave, New York, NY 10010, USA}
\pagerange{\pageref{firstpage}--\pageref{LastPage}} \pubyear{2017}
\date{30 January 2017}

\begin{document}
\maketitle
\label{firstpage}
\begin{abstract}
  The spatial variations of the velocity field of local stars provide
  direct evidence of Galactic differential rotation. The local
  divergence, shear, and vorticity of the velocity field---the
  traditional Oort constants---can be measured based purely on
  astrometric measurements and in particular depend linearly on proper
  motion and parallax. I use data for 304,267 main-sequence stars from
  the \emph{Gaia} DR1 Tycho-\emph{Gaia} Astrometric Solution to
  perform a local, precise measurement of the Oort constants at a
  typical heliocentric distance of 230 pc. The pattern of proper
  motions for these stars clearly displays the expected effects from
  differential rotation. I measure the Oort constants to be: $A =
  15.3\pm0.4\kms\kpc\inv$, $B = -11.9\pm0.4\kms\kpc\inv$, $C =
  -3.2\pm0.4\kms\kpc\inv$ and $K = -3.3\pm0.6\kms\kpc\inv$, with no
  color trend over a wide range of stellar populations. These first
  confident measurements of $C$ and $K$ clearly demonstrate the
  importance of non-axisymmetry for the velocity field of local stars
  and they provide strong constraints on non-axisymmetric models of
  the Milky Way.
\end{abstract}

\section{Introduction}

Determining the rotation of the Milky Way's disk is difficult, because
our vantage point is corotating with the local stars. As first pointed
out by \citet{Oort27a} and generalized by \citet{Ogrodnikoff32a}, the
local rotational frequency and the local change in the circular
velocity can be determined from the pattern of line-of-sight
velocities and proper motions of nearby stars as a function of
Galactic longitude $l$. Expanding the Milky Way's in-plane velocity
field to first order in heliocentric distance in an axisymmetric
velocity field, the azimuthal shear and vorticity contribute terms
proportional to $\cos 2l$, $\sin 2l$, and a constant to the mean
proper-motion and line-of-sight velocity, which can be distinguished
from the $\cos l$ and $\sin l$ pattern due to the Sun's peculiar
motion with respect to nearby stars. Going beyond the axisymmetric
approximation, the radial shear and divergence contribute similar
$\cos 2l$, $\sin 2l$, and constant terms. The four first-order terms
(azimuthal and radial shear, vorticity, and divergence) are known as
the Oort constants. Measurements of these constants provided the first
strong evidence that the Milky Way is rotating differentially with a
close-to-flat rotation curve \citep{Oort27a}.

The current best measurements of the Oort constants use Cepheids to
investigate the velocity field on large scales ($>1\kpc$) with a
kinematically-cold stellar tracer population
\citep[\eg,][]{Feast97a,Metzger98a}. Because the relative contribution
to the velocity pattern from the Sun's peculiar motion to that from
Galactic rotation diminishes as the inverse of the distance, the
intrinsic velocity field can be more easily determined from such
large-scale observations. However, higher-order contributions to the
velocity field become important at large distances and the derived
values for the Oort constants may not reflect their local value. Local
measurements require large stellar samples with good velocity
measurements. \citet{Olling03a} attempted a local measurement using
proper motions from the Tycho-2 catalog \citep{Hog00a} that was
complicated by the unknown distance distribution at different $l$,
which produces spurious terms in the velocity pattern that are
difficult to distinguish from the effects of Galactic rotation.

The \emph{Gaia} mission \citep{GaiaMission} has recently released its
first set of data, including parallaxes and proper motions for about 2
million Tycho-2 stars in the Tycho-$Gaia$ Astrometric Solution
\citep[TGAS,][]{Michalik15a,GaiaDR1,Lindegren16a}. This large set of
astrometric measurements allow the first truly local precise
measurement of the Oort constants. I discuss the definition and
interpretation of the Oort constants in
\sectionname~\ref{sec:method}. I describe the data used in this paper
in detail in \sectionname~\ref{sec:data} and the measurement of the
Oort constants in \sectionname~\ref{sec:oort}. I discuss the results
in \sectionname~\ref{sec:conclusion}.

\section{Definitions}\label{sec:method}

{\allowdisplaybreaks The four Oort constants $A$, $B$, $C$, and $K$
  are a representation of the four first-order expansion coefficients
  of the two-dimensional, in-plane mean velocity field of a stellar
  population in a Taylor series expansion of the mean velocity field
  with respect to distance from the Sun. For a given stellar
  population, in Galactocentric cylindrical coordinates $(R,\phi)$
  with the Sun at $(R_0,0)$ they are given by
\begin{align*}
2A & = \phantom{-}\bar{v}_\phi/R_0 - \bar{v}_{\phi,R}-\bar{v}_{R,\phi}/R_0\,,\\
2B & = -\bar{v}_\phi/R_0- \bar{v}_{\phi,R}+\bar{v}_{R,\phi}/R_0\,,\\
2C & = -\bar{v}_R/R_0+ \bar{v}_{R,R}-\bar{v}_{\phi,\phi}/R_0\,,\\
2K & = \phantom{-}\bar{v}_R/R_0+ \bar{v}_{R,R}+\bar{v}_{\phi,\phi}/R_0\,,
\end{align*}
where subscripts $,R$ and $,\phi$ denote derivatives with respect to
$R$ and $\phi$, respectively, and these derivatives are evaluated at
the Sun's position. In these expressions, $(\bar{v}_\phi,\bar{v}_R)$
is the mean rotational and radial velocity, respectively, of the
stellar population. The proper motion $(\mu_l,\mu_b)$ of the stellar
population can be expressed in terms of these constants as
\begin{equation}\label{eq:pml}
\begin{split}
\mu_l (l,b,\varpi) = & (A\,\cos 2l - C\,\sin 2l + B)\,\cos b \\
& \qquad + \varpi\,(u_0\,\sin l - v_0\,\cos l)\,,
\end{split}
\end{equation}
and
\begin{equation}\label{eq:pmb}
\begin{split}
\mu_b (l,b,\varpi) = & -(A\,\sin 2l + C\,\cos 2l + K)\,\sin b\,\cos b \\ &
\qquad + \varpi\,\left[(u_0\,\cos l + v_0\,\sin l)\,\sin b - w_0\,\cos
  b\right]\,,
\end{split}
\end{equation}
where $(u_0,v_0,w_0)$ is the Sun's motion with respect to the stellar
population and $\varpi$ is the inverse distance to the stars. See
\citet{Olling03a} for an elegant derivation of these relations.}

In the discussion above the term `Oort constants' is a misnomer
because (a) they vary for different stellar populations and (b) even
for a given stellar population they are not constant in
time. Historically, the Oort constants have been defined based on a
population of stars on (hypothetical) closed orbits, for which a
measurement of the Oort constants provides direct constraints on the
gravitational potential (which directly sets the properties of closed
orbits). Under the further assumption that the Galaxy is axisymmetric,
$C = K = 0$ and $\bar{v}_\phi = V_c=R\,\partial \Phi / \partial R$,
the circular velocity for the axisymmetric potential $\Phi$, and $A$
and $B$ then provide direct measurements of the angular rotation
frequency $\Omega_0 = V_c/R_0$ at the Sun and the local slope of the
rotation curve $\dd V_c/\dd R$.

The real Galaxy, however, is not this simple. Firstly, all stars have
attained a random component to their orbital energy that causes their
orbits to be eccentric and non-closed, even in an axisymmetric
potential. Thus, their mean velocity field is not related in a simple
manner to the gravitational potential. Secondly, non-axisymmetric or
time-dependent perturbations to the potential complicate the orbits of
even the coldest stellar populations and they typically no longer
close. The first effect can in principle be accounted for in a
relatively straightforward manner for well-mixed stellar populations
in an axisymmetric potential \citep{Bovy15a}, but in the likely
situation that the second effect is relevant, no precise, general
relation between the Oort constants and the potential of the Milky Way
exist. In this case, the measured values of the `Oort constants'
merely provide a strong constraint on the potential.

\section{Data}\label{sec:data}

\begin{figure}
  \includegraphics[width=0.48\textwidth,clip=]{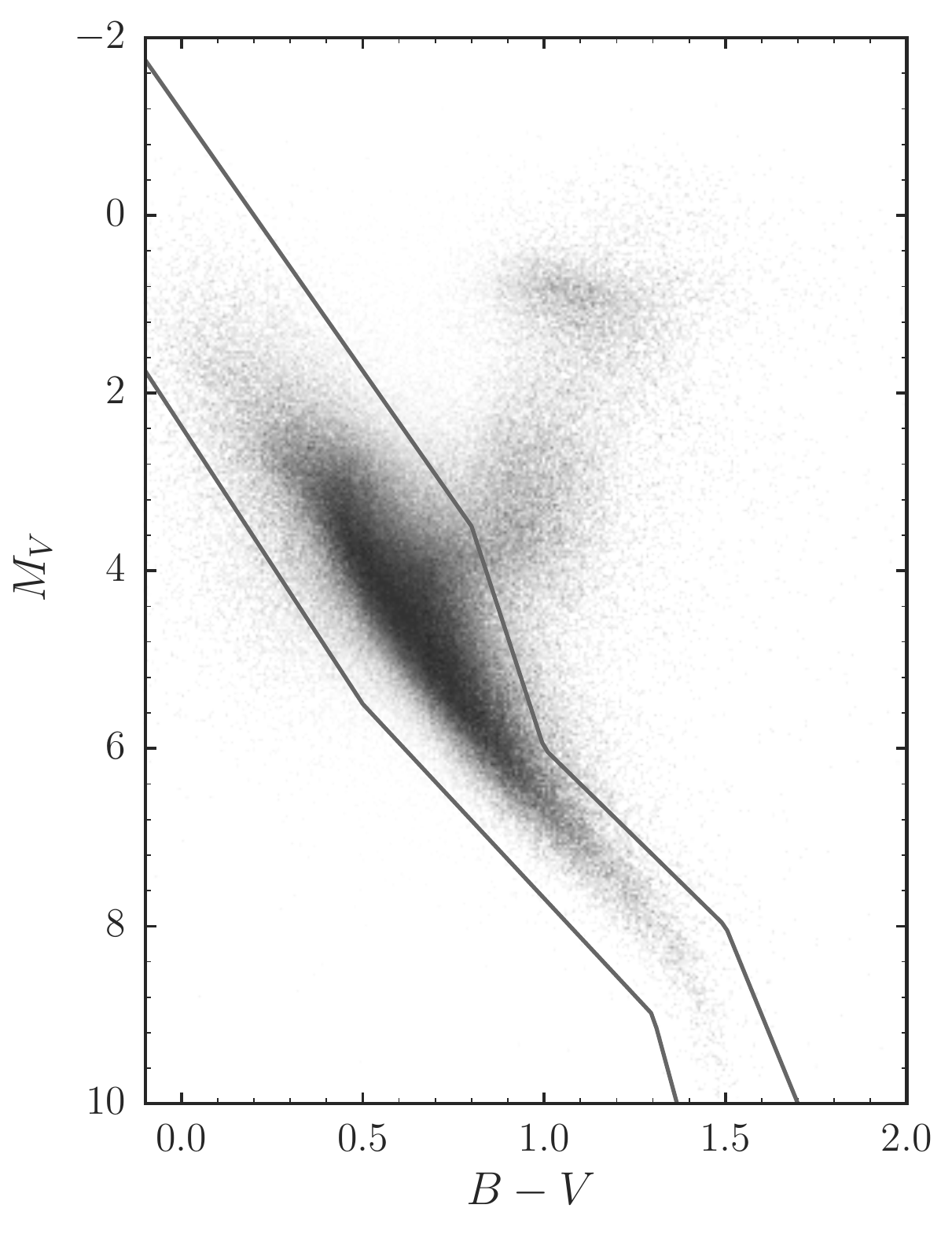}
  \caption{Color--magnitude diagram for 355,753 TGAS stars with
    relative parallax uncertainties less than 10\,\% within 500 pc
    that have $(B,V)$ photometry from APASS. The gray curves indicate
    the cuts that are employed to select the 315,946 main-sequence
    stars that are used in this paper.}\label{fig:cmd}
\end{figure}

\begin{figure}
  \includegraphics[width=0.42\textwidth,clip=]{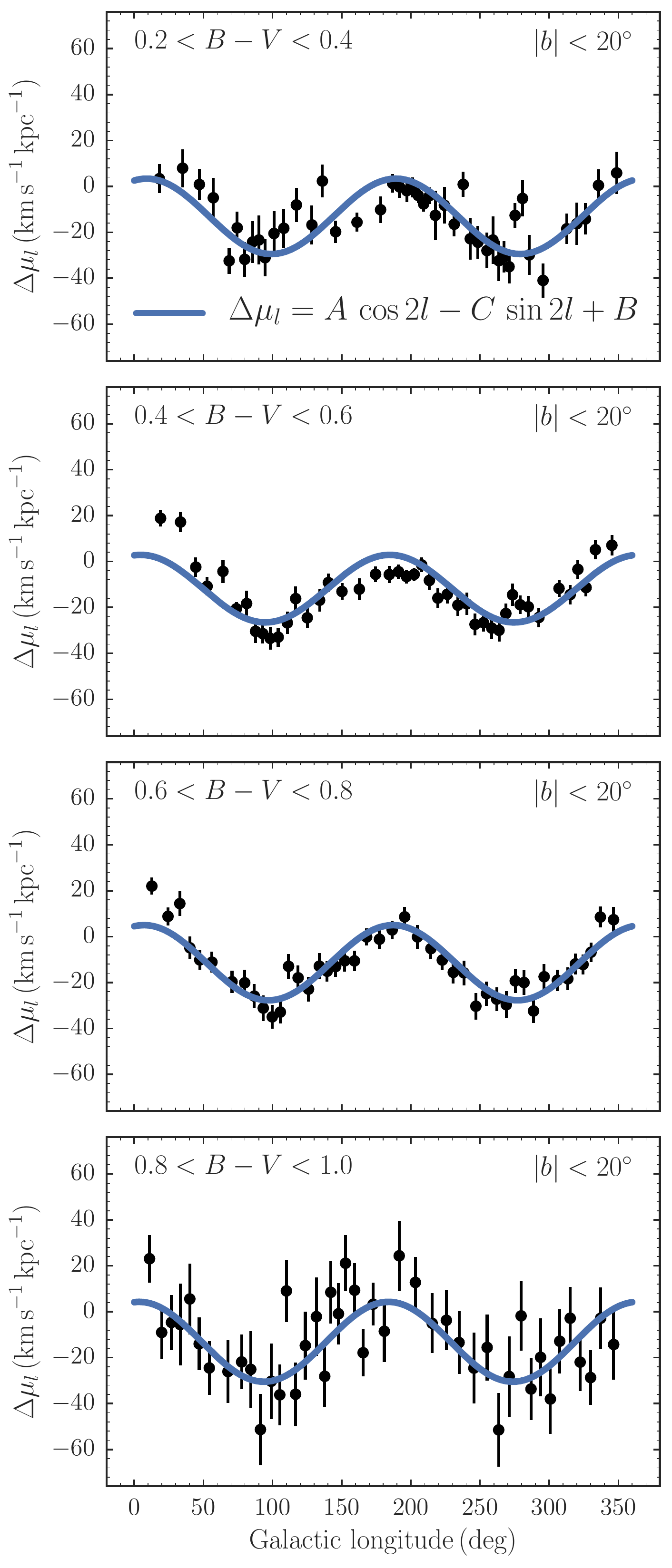}
  \caption{Comparison between the observed mean proper motion in
    Galactic longitude corrected for the solar motion (see
    \equationname s~[\ref{eq:dpml}]) as a function of $l$ and the
    best-fit model for the four main color bins used in the
    analysis. The data clearly display the expected signatures due to
    the differential rotation of the Galactic disk. The agreement
    between the model and the data is good.}\label{fig:pml}
\end{figure}

\begin{figure}
  \includegraphics[width=0.42\textwidth,clip=]{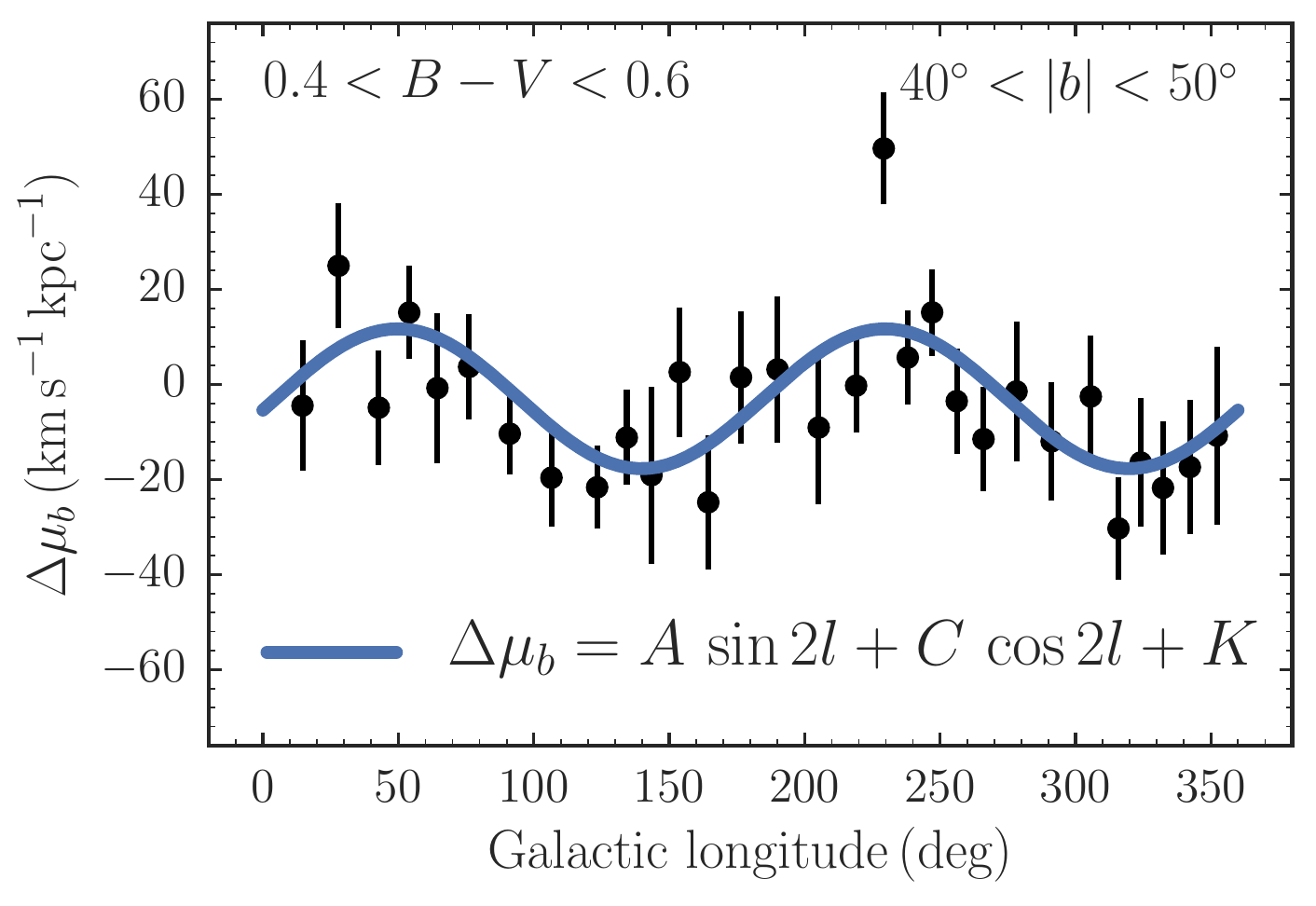}
  \caption{Same as \figurename~\ref{fig:pml}, but for the proper
    motion in Galactic latitude for a single color bin with many stars
    at the intermediate latitudes where Galactic rotation is
    clearest. The proper motion is corrected for the solar motion (see
    \equationname s~[\ref{eq:dpmb}]). The data displays the expected
    $\sin 2l$ behavior due to differential rotation and the non-zero
    mean offset shows that the local divergence is non-zero due to
    non-axisymmetric motions.}\label{fig:pmb}
\end{figure}

I employ parallaxes and proper motions from the \emph{Gaia} DR1 TGAS
catalog, which have typical uncertainties of $0.3$\,mas and
$1$\,mas\,yr$^{-1}$, respectively. I only consider the 450,278 TGAS
sources with relative parallax uncertainties less than $10\,\%$ and
with inverse parallaxes less than 500 pc to select a local sample. In
order to separate stellar populations into kinematically colder and
warmer populations, I use matched photometry from APASS
\citep{Henden12a} for the 355,753 stars with APASS photometry to
construct the $(B-V,M_V)$ color--magnitude diagram displayed in
\figurename~\ref{fig:cmd}. I select main-sequence stars using the cuts
shown in this figure---simply chosen by hand to encompass the
main sequence with minimal contamination---which creates a sample of
315,946 stars.

For this sample, I convert the proper motions in equatorial
coordinates to Galactic coordinates and propagate the proper-motion
uncertainty covariance matrix through this transformation as
well. Proper motions in mas\,yr$^{-1}$ are converted to units of
$\kms\kpc\inv$ by multiplying them by $4.74047$. In what follows, I
ignore the correlations between the measurements of the parallax and
the Galactic proper motions, because the scatter in the proper motions
is dominated by intrinsic scatter (the intrinsic scatter is
$\approx150\kms\kpc\inv$ due to the small distance range of the sample
versus typical measurement uncertainties
$\lesssim5\kms\kpc\inv$). Decorrelating the parallax and proper motion
errors by adding random, uncorrelated noise that is four times larger
than the formal uncertainties gives changes in the inferred Oort
constants below $\lesssim0.2\kms\kpc\inv$, confirming that the
parallax--proper-motion correlation is a subdominant source of
uncertainty.

\section{Measurement}\label{sec:oort}

\begin{figure*}
  \includegraphics[width=0.99\textwidth,clip=]{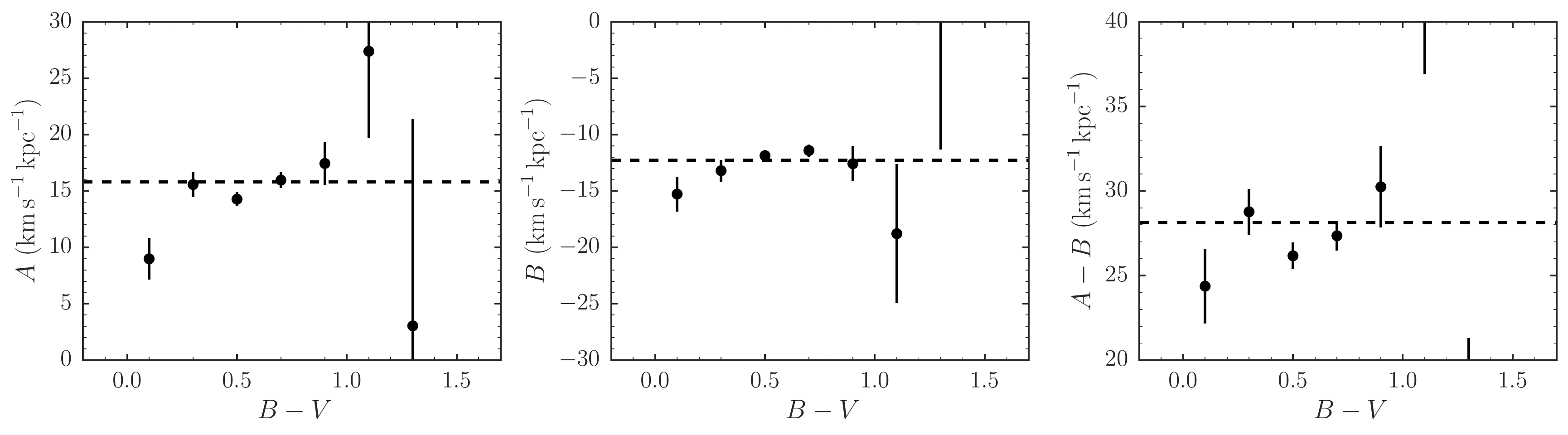}
  \caption{Measurements of $A$, $B$, and $A-B$ from the TGAS sample
    for different color bins. The measured values from different color
    bins agree well, except for the bluest, youngest stars which may
    be affected by streaming motions at birth. The dashed line in each
    panel indicates the combined measurement from the four color bins
    blueward of $B-V = 1$ (except for the bluest bin).}\label{fig:AB}
\end{figure*}

It is clear from the expressions in \equationname s~(\ref{eq:pml}) and
(\ref{eq:pmb}) that the Oort constants can be measured purely based on
astrometric quantities $(l,b,\varpi,\mu_l,\mu_b)$. The astrometric
quantities $(\varpi,\mu_l,\mu_b)$ appear in a linear fashion and are
therefore well behaved. Specifically, it is unnecessary to invert the
parallax to obtain a distance for the measurement of the Oort
constants. I model the observed individual proper motions using the
model in \equationname s~(\ref{eq:pml}) and (\ref{eq:pmb}), adding
intrinsic Gaussian scatter to account for the non-zero velocity
dispersion. The free parameters are therefore $(A,B,C,K,u_0,v_0,w_0)$
and two free parameters $(\sigma_{\mu_l},\sigma_{\mu_b})$ describing
the Gaussian scatter in $(\mu_l,\mu_b)$\footnote{For simplicity, the
  scatter is assumed to be independent of $(l,b,\varpi)$. This
  assumption is incorrect in detail, because the velocity ellipsoid in
  the Solar neighborhood is triaxial
  \citep[\eg,][]{Dehnen98a}. However, it does not negatively impact
  the results obtained here, because it will only cause the inferred
  scatter to be too large at some $(l,b,\varpi)$. This does not bias
  the best-fit value of the parameters $(A,B,C,K,u_0,v_0,w_0)$
  describing the mean trend of $\mu_l(l)$ and $\mu_b(l)$ given in
  \equationname s~(\ref{eq:pml}) and (\ref{eq:pmb}), but it will
  slightly overestimate the uncertainties in these values.}.

I fit the data $(\mu^i_l,\mu^i_b)$ with associated uncertainties
$(s^i_{\mu_l},s^i_{\mu_b})$ sliced in $0.2$\,mag bins
in $B-V$ using the log likelihood
\begin{equation*}
\begin{split}
  \ln \mathcal{L} = -\frac{1}{2}\,\sum_i\,\Bigg(& \frac{[\mu^i_{l}-\mu_l(l_i,b_i,\varpi_i)]^2}{\sigma_{\mu_l}^2+(s^i_{\mu_l})^2}+\ln[\sigma_{\mu_l}^2+(s^i_{\mu_l})^2]\\  + & \frac{[\mu^i_{b}-\mu_b(l_i,b_i,\varpi_i)]^2}{\sigma_{\mu_b}^2+(s^i_{\mu_b})^2}+\ln[\sigma_{\mu_b}^2+(s^i_{\mu_b})^2]\Bigg)\,,
\end{split}
\end{equation*}
The parameters $A,C,u_0$ and $v_0$ are constrained by both components
of the proper motions; fitting them separately to $\mu_l$ and $\mu_b$
gives consistent results for all four of these. I determine the
uncertainty in the best-fit parameters using Markov Chain Monte Carlo
with flat priors on all parameters. Note that all parameters except
the scatter terms are linear parameters whose probability distribution
at fixed scatter could be obtained analytically, but because we need
to fit for the non-linear scatter parameters I do not make use of this
simplification.

The best-fitting model to four color bins is compared to the $\mu_l$
data in \figurename~\ref{fig:pml}. Because the solar motion is the
dominant effect on $\mu_l(l)$ that obscures the Galactic rotation
contribution, the observed proper motions in this figure have been
corrected for the solar motion $(u_0,v_0)$ using the best-fit
parameters as follows
\begin{equation}\label{eq:dpml}
\Delta \mu_l (l) = \left(\mu_l - \varpi\,[u_0\,\sin l + v_0\,\cos l]\right)\big/\cos b\,,
\end{equation}
such that the mean $\Delta \mu_l(l) = A\,\cos 2l - C\,\sin 2l +
B$. The data are binned to display the mean $\Delta \mu_l(l)$ because
of the large scatter of individual data points. On account of the
$1/\cos b$ correction, in \figurename~\ref{fig:pml} I only show stars
with $|b| < 20^\circ$ that display the trend most clearly, even though
stars at all $b$ are used in the fit.

The proper motion in Galactic latitude is much more tenuously related
to the parameters of Galactic rotation on account of the $\sin b\,\cos
b$ factor in \equationname~(\ref{eq:pmb}). Therefore, I only compare
the data and the model in \figurename~\ref{fig:pmb} for the single
$B-V$ bin that contains enough stars at intermediate $b$ to allow for
a straightforward data--model comparison. Similar to
\figurename~\ref{fig:pml}, the proper motion in
\figurename~\ref{fig:pmb} is corrected for the solar motion as follows
\begin{align}\label{eq:dpmb}
& \Delta \mu_b (l) = \\
& \!\!\!\!\!-\left(\mu_b - \varpi\,\left[(u_0\,\cos l + v_0\,\sin l)\,\sin b - w_0\,\cos b\right]\right)\big/(\sin b \cos b)\,,\nonumber
\end{align}
such that the mean $\Delta \mu_b(l) = A\,\sin 2l + C\,\cos 2l +
K$. The data are again binned to display the mean trend and I restrict
the sample to stars with $40^\circ < |b| < 50^\circ$, even though
stars at all $b$ are used in the fit.

Overall the data display good agreement with the simple first-order
velocity model in both $\mu_l(l)$ and $\mu_b(l)$. The posterior
probability distribution for the Oort and solar-motion parameters
displays no correlations among any of the parameters, as expected from
the full and close to uniform coverage in $l$ of the stars in the TGAS
sample.

\figurename~\ref{fig:AB} shows the measurements of $A$, $B$, and
$A-B$, which is equal to the rotational frequency $\Omega_0$ if the
Galaxy were axisymmetric, as a function of $B-V$. The different color
bins display excellent agreement with each other, except for the very
bluest stars, which are likely affect by residual streaming motions
from birth. That $A$ and $B$ do not strongly depend on color (and,
thus, on velocity dispersion) is expected in the limit of well-mixed,
solar-neighborhood populations in an axisymmetric Galaxy, for which
$A$ and $B$ should be constant to within $\lesssim 1\kms\kpc\inv$
\citep{Bovy15a}. I combine the four bins blueward of $B-V = 1$ (except
for the bluest bin) which contain the majority of the stars in the
sample (304,267 out of 315,946) to obtain a single value of each Oort
constant. The combined measurements are: $A = 15.3\pm0.4\kms\kpc\inv$,
$B = -11.9\pm0.4\kms\kpc\inv$ as well as $A-B =
27.1\pm0.5\kms\kpc\inv$ and $-(A+B) = -3.4\pm0.6\kms\kpc\inv$.

\begin{figure*}
  \includegraphics[width=0.99\textwidth,clip=]{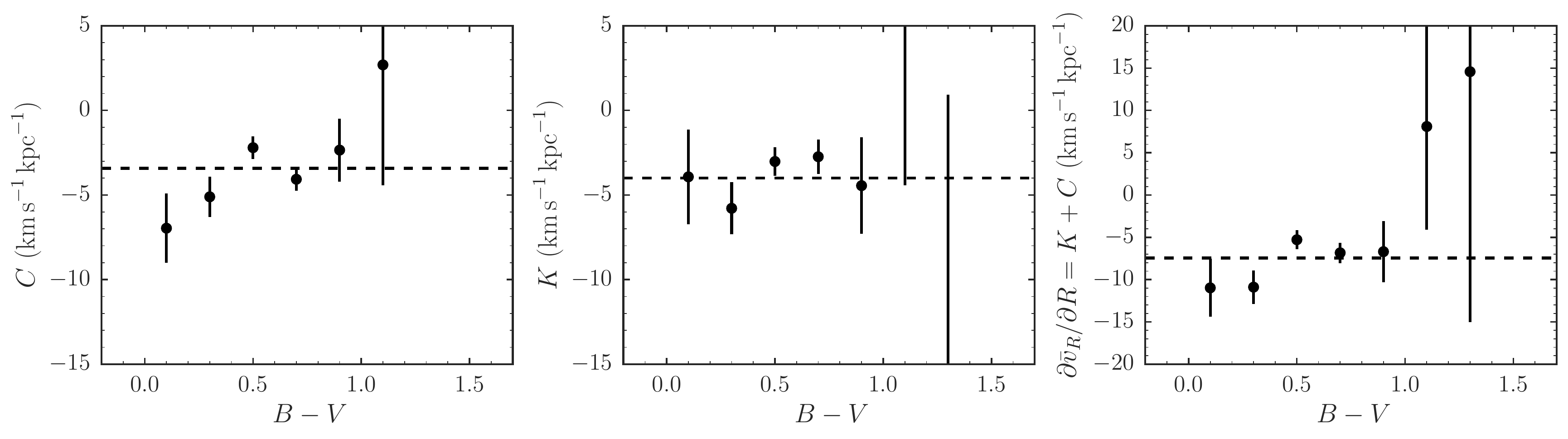}
  \caption{Measurements of $C$, $K$, and $K+C$ from the TGAS sample
    for different color bins. The values derived from all of the
    different color bins show good agreement, with no hint of a trend
    with color in any of the measurements. The dashed line in each
    panel indicates the combined measurement from the four color bins
    blueward of $B-V = 1$ (except for the bluest bin).}\label{fig:CK}
\end{figure*}

The measurements of $C$ and $K$ are displayed in
\figurename~\ref{fig:CK}. Like for $A$ and $B$, the measurements of
$C$ and $K$ in different color bins agree with each other. The
measurements are: $C = -3.2\pm0.4\kms\kpc\inv$ and $K =
-3.3\pm0.6\kms\kpc\inv$. The reason that the uncertainty in $K$ is
larger than that in the other parameters is that it can only be
determined from the behavior of $\mu_b(l)$. I also find that $K+C=
\frac{\partial \bar{v}_R}{\partial R}\big|_{R_0} =
-6.6\pm0.7\kms\kpc\inv$. The best-fitting $C$ and $K$ are therefore
both significantly non-zero, indicating that the local kinematics is
non-axisymmetric. There is no trend in color for either $C$ or $K$, in
disagreement with the previous measurements from \citet{Olling03a}.

\section{Discussion and conclusion}\label{sec:conclusion}

I have determined the Oort constants using astrometric measurements
for 304,267 local (typical distance of 230\pc) main-sequence stars
from the \emph{Gaia} DR1 TGAS catalog: {\allowdisplaybreaks
\begin{align}
  A & = \phantom{-}15.3\pm0.4\kms\kpc\inv\,,\\
  B & = -11.9\pm0.4\kms\kpc\inv\,,\\
  C & =  -3.2\pm0.4\kms\kpc\inv\,,\\
  K & = -3.3\pm0.6\kms\kpc\inv
\end{align}}

These measurements for $A$ and $B$ are in good agreement with those
obtained from modeling the Hipparcos proper motions of Cepheids: $A =
14.82 \pm 0.84\kms\kpc\inv$ and $B= -12.37 \pm 0.64\kms\kpc\inv$
\citep{Feast97a}. The measurement of $A$ is also in good agreement
with that measured from the line-of-sight velocities of Cepheids: $A =
15.5\pm0.4\pm1.2\,(\mathrm{syst.})\kms\kpc\inv$
\citep{Metzger98a}. Unlike these measurements, the current
measurements are based on a sample of stars within a few 100 pc from
the Sun and thus truly measure the local velocity field.

The TGAS data also allow a confident measurement of $C$ and $K$. There
are only a few previous measurements of these
constants. \citet{Olling03a} measured $C$ from the Tycho-2 proper
motions, correcting for the influence of the unknown distance
distribution, and found $C = -10\pm2\kms\kpc\inv$ and that $C$ is more
negative for older, kinematically-warmer populations. We find that $C
= -3.2\pm0.4\kms\kpc\inv$ with no dependence on the color or velocity
dispersion of the stellar population. Previous measurements of $K$
using young stars find that it is negative with typical values in the
range $-3\kms\kpc\inv$ to $-1\kms\kpc\inv$ \citep{Comeron94a,Torra00a}
in agreement with the present measurement. The measurement of $K+C=
\frac{\partial \bar{v}_R}{\partial R}\big|_{R_0} =
-6.6\pm0.7\kms\kpc\inv$ is in good agreement with the measurement of
the local mean-radial-velocity gradient from RAVE \citep{Siebert11a}.

That $C$ and $K$ are both non-zero for all color-selected stellar
populations provides strong evidence that the local velocity field is
shaped by non-axisymmetric forces. The localized nature of the
measurements using the TGAS sample makes these measurements
straightforward constraints on any large-scale non-axisymmetric model
such as the bar or spiral arms, in which the Oort constants may simply
be evaluated at the position of the Sun and compared to these
measurements. For example, for no plausible model with a triaxial
bulge or halo in which closed orbits are elliptical do $C$ and $K$
have the same sign \citep{Kuijken94a,Bovy15a}, such that this model on
its own is inconsistent with the current measurements. For an
axisymmetric disk plus bar model with the bar's outer Lindblad
resonance near the Sun---which can explain the Hercules moving group
in the local velocity distribution \citep{Dehnen00a} and the power
spectrum of velocity fluctuations in the disk \citep{Bovy15b}---$C$
and $K$ are both negative with $C \approx K \approx -2\kms\kpc\inv$
(computed using \texttt{galpy}; \citealt{Bovy15a}), although
kinematically-colder populations in this model respond differently to
the bar perturbation and give different $C$ \citep{Minchev07a}, which
is not observed in the TGAS data. More sophisticated modeling is
required to better interpret the measurements of the Oort constants
presented here.

That $|C| \approx |K| \approx 0.2 \times |A|$ means that non-circular
streaming motions in the solar neighborhood are important and that $A$
and $B$ cannot be directly related to the Milky Way's circular
velocity and its slope at the Sun, as is commonly assumed (see
\sectionname~\ref{sec:method}). A simple estimate of the difference
between the value of the circular velocity derived from local
measurements of $A$ and $B$ (which is $[A-B]\,R_0 \approx 220\kms$
using the measured $A$ and $B$ and $R_0\approx8\kpc$) and the global,
azimuthally-averaged value is $(|C|+|K|)/(|A|+|B|)\approx 20\,\%$ or
$\approx40\kms$ for a circular velocity $\approx 220\kms$. This result
more generally applies to determinations of the circular velocity from
the local kinematics of stars or gas, which are affected by similar
streaming motions.

{\bf Acknowledgements} I thank Dustin Lang for providing the
TGAS-matched APASS data, the 2016 NYC Gaia Sprint participants for
stimulating discussions, and the anonymous referee for a helpful
report. I also thank Maarten Breddels for finding and fixing a bug in
\figurename s~\ref{fig:pml} and \ref{fig:pmb}. JB received support
from the Natural Sciences and Engineering Research Council of
Canada. JB also received partial support from an Alfred P. Sloan
Fellowship and from the Simons Foundation. The MCMC analyses in this
work were run using \emph{emcee} \citep{Foreman13a}.

This project was developed in part at the 2016 NYC Gaia Sprint, hosted
by the Center for Computational Astrophysics at the Simons Foundation
in New York City.

This work has made use of data from the European Space Agency (ESA)
mission {\it Gaia} (\url{http://www.cosmos.esa.int/gaia}), processed
by the {\it Gaia} Data Processing and Analysis Consortium (DPAC,
\mbox{\url{http://www.cosmos.esa.int/web/gaia/dpac/consortium}}). Funding for
the DPAC has been provided by national institutions, in particular the
institutions participating in the {\it Gaia} Multilateral
Agreement. This research was made possible through the use of the
AAVSO Photometric All-Sky Survey (APASS), funded by the Robert Martin
Ayers Sciences Fund.

\end{document}